\documentclass[aps,twocolumn,nofootinbib]{revtex4}
\usepackage[utf8]{inputenc}
\usepackage{epsfig}
\usepackage{amsfonts}
\usepackage{amssymb}
\usepackage{calrsfs}
\usepackage{amsmath}
\newcommand{\beq}{\begin{equation}}
\newcommand{\eeq}{\end{equation}}

\begin{document}

\title{General linear thermodynamics for periodically driven systems 
with multiple reservoirs}
\author{Karel Proesmans$^{1}$ and Carlos E. Fiore$^{2}$}
\address{$^{1}$ Hasselt University, B-3590 Diepenbeek, Belgium,\\
$^{2}$ Institute of Physics of S\~ao Paulo University, 
Rua do Mat\~ao, 1371, 05508-090
S\~ao Paulo, SP, Brazil}
\date{\today}

\begin{abstract}
We derive a linear thermodynamics theory 
for general Markov dynamics with both steady-state and 
time-periodic drivings.
Expressions for thermodynamic quantities, such as mechanical and
chemical work, heat and entropy production are obtained in terms
of equilibrium probability distribution and the drivings.
The entropy production is derived as a bilinear function of
thermodynamic forces and the associated fluxes. We derive explicit
formulae for the Onsager coefficients and use them to verify
the Onsager-Casimir reciprocal relations. 
Our results are illustrated on a periodically driven quantum dot in contact with two electron reservoirs and optimization protocols are discussed.
\end{abstract}

\maketitle

\section{Introduction}
Due to the seminal work of primarily Onsager and Prigogine, the theory of linear irreversible thermodynamics has become one of the cornerstones of modern statistical physics. Close to equilibrium, one can use this framework to determine the thermodynamic fluxes, such as heat and work, and show that they satisfy general properties, such as Onsager symmetry and the Green-Kubo relations \cite{de2013non}.

Over the last two decades, a somewhat different approach has been
undertaken to study the thermodynamics of small-scaled
systems \cite{seifert2012stochastic,van2015ensemble}.
This theory, known as stochastic thermodynamics, uses
Markov dynamics to model systems at the mesoscale, where fluctuations in the thermodynamic fluxes become important. The assumption of
local detailed balance then leads to a consistent definition of
the thermodynamic properties of the system. The stochastic fluxes
of the system satisfy general relations such as the
Jarzynski equality \cite{jarzynski1997nonequilibrium,saira2012test}.
Furthermore, this theory has lead to applications in several other branches of
science, such as information theory \cite{parrondo2015thermodynamics},
chemical reaction networks \cite{rao2016nonequilibrium}, and active
matter \cite{mandal2017entropy,dabelow2019irreversibility}.

A natural question to ask is how the classical ideas of 
linear irreversible thermodynamics can be incorporated in the  theory of stochastic thermodynamics. 
This problem has been addressed for several 
case-studies \cite{izumida2009onsager,izumida2010onsager,rosas2016onsager2,rosas2016onsager,proesmans2016brownian,cerino2016linear,brandner2016periodic,yamamoto2016linear,proesmans2017underdamped,rosas2017stochastic}. Furthermore, 
general theories have been derived for periodically 
driven systems in contact with a single reservoir \cite{brandner2015thermodynamics,proesmans2015onsager,proesmans2016linear}, and
for steady-state systems in contact with two
reservoirs \cite{tome2012entropy,tome2015stochastic}, leading to bounds the power and efficiency of thermodynamic engines \cite{bauer2016optimal,proesmans2016power,iyyappan2018relations}.
A  more general approach for systems with any number of reservoirs and time-dependent driving has not been studied thoroughly.

In this paper, we close this gap by deriving a general formalism for
the linear thermodynamics of stochastic systems with both
steady-state and time-periodic drivings. Our study
is carried out by taking into account multiple heat and
particle reservoirs. We obtain expressions for
thermodynamic quantities, such as mechanical and
chemical work, heat and entropy production in terms
of equilibrium probability distribution and the drivings. In particular, we show that general results of linear irreversible thermodynamics, such as the structure of entropy production rate and Onsager symmetry are valid for this general class of systems.

This paper is organized as follows.
We start in section \ref{mod} by introducing the model and by discussing its linearized dynamics. In section \ref{tf}, we define the work and heat fluxes and show how they are related to the entropy production rate. The evaluation of Onsager coefficients  and the existence of an Onsager-Casimir symmetry relation are discussed in \ref{ons}. In section \ref{app}, we apply our formalism to a periodically driven two-level 
system. Conclusions and outlook are discussed in section \ref{con}.

\section{Model \label{mod}}
Throughout this paper, we focus on systems with a discrete set of states
in contact with multiple temperature and particle reservoirs
that can induce transitions between distinct configurations.
The system can be in a given state $m$, specified  by its energy 
$\epsilon_m(t)$ and particle number $n_m$, with probability $p_m(t)$. The time-evolution of $p_m(t)$ is described by a master equation
\begin{equation}
 {\dot p}_{m}(t)=\sum_{n,j}W^{j}_{mn}(t)p_n(t),
\label{eq2}
\end{equation}
where $W^{j}_{mn}(t)$ is  the probability  per unit of time of a 
transition from a state $n$ to a state $m$ induced
by reservoir $j$.
Conservation of probability implies
that
\begin{equation}
\sum_{m}W^{j}_{mn}(t)=0,
\label{eq2-1}
\end{equation}
valid for all $m$, and therefore
$W^{j}_{mm}(t)=-\sum_{n \neq m}W^{j}_{nm}(t)$.

Each reservoir $j$ is characterized by a temperature $T_j(t)$ and
chemical potential $\mu_j(t)$.
If the system is in contact with a single reservoir 
with time-independent temperature $T_{j}$ and chemical potential $\mu_{j}$,
it will converge to
an equilibrium state given by the Boltzmann-Gibbs
distribution:
\begin{equation}
  P^{j}_m=\frac{1}{Z_{\rm eq}^{j}}e^\frac{-(\epsilon_{m}-\mu_{j}n_m)}{T_{j}},
  \end{equation}
where $Z_{\rm eq}^j=\sum_{m}e^{-(\epsilon_{m}-\mu_{j}n_m)/T_{j}}$ is
the (grand-canonical) partition function.
By definition, the above equilibrium distribution should satisfy
the detailed balance
condition, $W^{j,eq}_{mn}P^{j}_n-W^{j,eq}_{nm}P^{j}_m=0$,
implying the following ratio between the transition
rates $W^{j,eq}_{mn}$ and $W^{j,eq}_{nm}$ 
\begin{equation}
\frac{W^{eq;j}_{mn}}{W^{eq;j}_{nm}}=e^{-\{(\epsilon_{m}-\epsilon_{n})-\mu_{j}(n_m-n_n)\}/T_{j}}.
\label{eq3} 
\end{equation}
This expression allows us to write the transition rate $W^{eq;j}_{mn}$
as follows:
\begin{equation}
W^{eq;j}_{mn}=C^{j}_{mn}\lambda^{eq;j}_{n},
\label{eq4}
\end{equation}
where $\lambda^{eq;j}_{n}=\exp\left({(\epsilon_{n}-\mu_{j}n_n)/T_{j}}\right)$
and  $C^{j}_{mn}$ is a matrix that quantifies the coupling strength 
between states $m$ and $n$. Due to the assumption of local detailed
balance and the properties of the transition matrix, $C^{j}$ satisfies
the following symmetry relations:
\begin{equation}
    C^{j}_{mn}=C^{j}_{nm}, \qquad C^{j}_{nn}=-\sum_{m\neq n}C^{j}_{mn}.\label{symC}
\end{equation}

\subsection{Linear description}
As stated before, the system will reach an equilibrium Boltzmann state when
it is in contact with a single reservoir at constant
temperature and chemical potential. This is
generally not the case when the system is in contact with multiple
reservoirs or when the temperature and chemical potential are
modulated time-periodically. In those cases, detailed balance is
broken, and the system starts  dissipating heat and producing entropy.
As each reservoir operates independently, { the}
transition rate of each reservoir 
has the same form as in  Eq.~(\ref{eq4}), but with time-dependent parameters 
$\epsilon_{m}(t),T_j(t),\mu_j(t)$ and $C^{j}_{mn}(t)$.
The total transition matrix is 
obtained by summing over all reservoirs, $W_{mn}(t)=\sum_{j}W^{j}_{mn}(t)$, where $W^{j}_{mn}(t)$
is given by Eq.~(\ref{eq4}) for every reservoir $j$.

The temperatures and chemical potentials are modulated time-periodically.
We introduce the driving functions $g_{T_j}(t)$ and
$g_{\mu_j}(t)$ as
\begin{eqnarray}
\frac{1}{T_j(t)}&=&\frac{1}{T_{0,j}}+F_{T_j}g_{T_j}(t),
\label{eq8}\\
\mu_j(t)&=&\mu_{0,j}+T_{0,j}F_{\mu_j}g_{\mu_j}(t),
\label{eq9}
\end{eqnarray}
where  $F_{\alpha_j}$'s correspond to the strength of the
thermodynamic  drivings $\alpha_j \in \{T_j,\mu_j\}$.
The energy of each microscopic state is also driven periodically by an external work source,
\begin{equation}
\epsilon_n(t)=\epsilon_{0,n}+T_{0,j}F_{\epsilon}\gamma_{\epsilon,n}g_{\epsilon}(t),
\label{eq7}
\end{equation}
where $\gamma_{\epsilon,n}$ is the amplitude
with which the level $n$ is modulated.
As we are focusing on the regime close to equilibrium, both temperature and chemical potential modulations are assumed to be
around the same equilibrium state 
for all reservoirs, $T_{0,j}=T_0$ and $\mu_{0,j}=\mu_0$ for all $j$. 

To make further progress, we assume that the thermodynamic
forces are sufficiently small so that  we can perform a linear
approximation. This is the crucial assumption for the theory of linear irreversible thermodynamics \cite{de2013non}.
By expanding  the coupling matrix
$C$ up to first-order with respect to  modulations of temperature, chemical potential and energy, we have that 
\begin{equation}
C^{j}_{mn}(t)=C^{eq,j}_{mn}+\sum_{\alpha,j}F_{\alpha_j}g_{\alpha_j}(t)C^{\alpha_j}_{mn}.
\label{eq10}
\end{equation}
The perturbed coupling matrices $C^{\alpha_j}$ should satisfy the same symmetry relations as the unperturbed coupling matrices, Eq.~(\ref{symC}).
They are used for
obtaining the following linear expression for $W^{j}_{mn}(t)$
in terms of thermodynamic forces:    
\begin{equation}
  W^{j}_{mn}(t)=W^{eq,j}_{mn}+\sum_{\alpha}F_{\alpha_j}W^{\alpha_j}_{mn}(t),
   \label{Walp-0}
\end{equation} 
where 
\begin{equation}W^{\alpha;j}_{mn}(t)=g_{\alpha_j}(t)[W^{eq,j}_{mn}\gamma_{n}^{\alpha}+C_{mn}^{\alpha_j}{\lambda^{eq;j}_{n}}],
  \label{Walp}
\end{equation}
and the vector $\gamma^{\alpha}$  has elements given by 
\begin{eqnarray}\gamma_{n}^{\alpha}=\begin{cases}(\epsilon_{0,n}-\mu_0n_n)& \alpha=T_j\\ \gamma_{\epsilon,n} & \alpha=\epsilon\\-n_n & \alpha=\mu_j.
\end{cases}
\end{eqnarray}

Since the driving functions $g_{\alpha_j}(t)$ are assumed to be time
periodic, $g_{\alpha_j}(t+\mathcal{T})=g_{\alpha_j}(t)$, with
$\mathcal{T}$ being the period of the driving,  the system will
relax to a time-periodic steady-state distribution.
This distribution can be expanded up to linear order in terms of the
thermodynamic forces \cite{proesmans2016linear},
\begin{equation}
p(t)=p^{eq}+\sum_{\alpha,j}F_{\alpha_j}p^{\alpha_j}(t),
\label{eq11} 
\end{equation}
where $p^{eq}$ is the Boltzmann-Gibbs distribution associated 
to the reference energy, temperature and chemical
potential, $\epsilon_{0,n},T_0$ and $\mu_0$, respectively.
Substituting Eq.~(\ref{eq11}) into the master equation (\ref{eq2}) leads to
\begin{equation}
{\dot p}^{\alpha_j}(t)=W^{eq}p^{\alpha_j}(t)+W^{\alpha_j}(t)p^{eq}.
\label{eq12} 
\end{equation}
Thus, the  first-order  contribution in the probability ${\dot p}^{\alpha_j}(t)$
depends only on the total equilibrium matrix $W^{eq}_{mn}=\sum_jW^{eq,j}_{mn}$
and on the linear perturbation matrix $W^{\alpha_j}(t)$
evaluated over the equilibrium probability.
The above expression can be integrated, leading to
\begin{equation}
p^{\alpha_j}(t)=\int_{0}^{\infty}{\rm d}\tau\, e^{W^{eq}\tau}W^{\alpha_j}(t-\tau)p^{eq},
\label{eq13} 
\end{equation}
which is time-periodic.
Plugging in the explicit formula for $W^{\alpha_j}(t)$  and once again taking
into account the
properties of the coupling matrix given by Eq.~(\ref{symC}), we arrive at
the final expression for the component $p^{\alpha_j}(t)$
\begin{equation}
p^{\alpha_j}(t)=\int_{0}^{\infty}{\rm d}\tau\, e^{W^{eq}\tau}W^{eq,j}\gamma^{\alpha}p^{eq}g_{\alpha_j}(t-\tau),\label{pj}
\end{equation}
i.e., the first order correction in the coupling matrix does not
lead to corrections in the probability distribution.
We conclude that, up to first-order,
the time-periodic steady-state distribution
can be obtained exactly in terms of the equilibrium properties,
the driving function $g_{\alpha}(t)$'s and the $\gamma_\alpha$'s.

Finally, it is worth mentioning that Eq.~(\ref{pj}) reduces to
the one obtained in Ref. \cite{proesmans2016linear} 
for the one reservoir case.

\section{Thermodynamic fluxes \label{tf}}
Having developed a general formalism to derive the linear dynamics of the system under study, we are now ready to evaluate the thermodynamic properties, using the framework of stochastic thermodynamics \cite{van2015ensemble,seifert2012stochastic}.
In particular, the mechanical work flux
${\dot W}_{mech}(t)$, chemical work flux ${\dot W_{chem}(t)}$ and 
the heat flux ${\dot Q(t)}$  are given by
\begin{eqnarray}
\dot{W}_{mech}(t)&=&\sum_{m}{\dot \epsilon_m(t)}p_m(t), 
\label{eq14-1}
\\
 {\dot W_{chem}(t)}&=&\sum_{m,j}\mu_j(t)n_m{\dot p^{j}_m(t)},
\label{eq14-2}
\\
{\dot Q(t)}&=&\sum_{m,j}[\epsilon_m-\mu_j(t)n_m]{\dot p^{j}_m(t)},
\label{eq14-3}
\end{eqnarray}
with
\begin{equation}
    \dot{p}^{j}_m(t)=\sum_n W^j_{mn}(t)p_n(t).
\end{equation}
The time evolution of the mean internal energy of the system
$U(t)=\sum_{m}\epsilon_m(t)p_m(t)$ is given by
\begin{equation}
{\dot U(t)}={\dot W_{mech}(t)}+{\dot W_{chem}(t)}+{\dot Q(t)},
\label{eq14}
\end{equation}
in agreement with the first law of thermodynamics.
By inserting Eq.~(\ref{eq7}) into Eq.~(\ref{eq14-1}),
the average mechanical work per unit of time can be written as
\begin{equation}
\dot{\bar{W}}_{mech}=\frac{T_0F_{\epsilon}}{\mathcal{T}}\int_{0}^{\mathcal{T}}dt\,{\dot g_{\epsilon}(t)}\sum_{m}\gamma_{\epsilon,m}p_m(t).
\label{eq15} 
\end{equation}
Eq.~(\ref{eq15}) is  conveniently rewritten 
 as a  product of forces and flux,
$\dot{\bar{W}}_{mech}=T_0F_{\epsilon}J_\epsilon$, with work flux
$J_\epsilon$ given by
\begin{equation}
J_\epsilon=-\frac{1}{\mathcal{T}}\int_{0}^{\mathcal{T}}dt\,{g_{\epsilon}(t)}\sum_{m} \gamma_{\epsilon,m}{\dot p}_m(t),
\label{eq15-1}
\end{equation}
where a partial integration was performed taking
into account the periodicity of $p_m(t)$.
One can decompose it further as
\begin{equation}
    J_{\epsilon}=\sum_{j}J_{\epsilon_j},
\end{equation}
with $J_{\epsilon_j}$ given by
\begin{equation}
    J_{\epsilon_j}=-\frac{1}{\mathcal{T}}\int_{0}^{\mathcal{T}}dt\,{g_{\epsilon}(t)}\sum_{m,n} \gamma_{\epsilon_j,m}W_{mn}^{j}(t)p_n(t) ,\label{Jepsi}
\end{equation}
and $T_0F_{\epsilon}J_{\epsilon_j}$
can be interpreted as the mechanical work delivered to reservoir $j$.

Proceeding analogously,
the total mean chemical work per cycle can be obtained by integrating Eq.~(\ref{eq14-2}) over one period and subtracting $\mu_0\sum_{j}\int_{0}^{\mathcal{T}}{\dot p}^{j}_m(t)dt=0$, which gives
\begin{equation}
\dot{\bar{W}}_{chem}=T_0\sum_jF_{\mu_j}J_{\mu_j},
\label{eq16} 
\end{equation}
where $J_{\mu_j}$ is defined as
\begin{eqnarray}
J_{\mu_j}&=&-\frac{1}{\mathcal{T}}\int_{0}^{\mathcal{T}}dt\,{g_{\mu_j}(t)}\sum_{m,n} \gamma_{\mu_j,m}W_{mn}^{j}(t)p_n(t).
\label{eq16-1} 
\end{eqnarray}

Since the driving is periodic, the  average
internal energy of the system
does not change over a period of the driving, ${\dot {\bar U}}=0$. The first law of thermodynamics then leads to  
an expression for the average heat  in terms of fluxes,
\begin{equation}
\dot{\bar{Q}}=-{\dot {\bar W}}-{\dot {\bar W}_{chem}}=-T_{0}\sum_j\left(F_{\epsilon_j} J_{\epsilon_j}+F_{\mu_j}J_{\mu_j}\right).
\label{eq17}
\end{equation}
This expression can also be  evaluated by a direct integration of 
Eq.~(\ref{eq14-3}) over one
period and summing over the contribution of all reservoirs.

The total entropy production $\bar {\sigma}$ is given by the sum of
the contribution of all reservoirs
\begin{equation}
\bar {\sigma}=\sum_j \bar {\sigma}^{j},
\end{equation}
where   each term $\bar {\sigma}^{j}$ can be calculated through
the microscopic formula \cite{schnakenberg1976network,jiu1984stability}
\begin{equation}
\bar {\sigma}^{j}=\frac{1}{\mathcal{T}}\sum_{m,n}\int_{0}^{\mathcal{T}}dt\,W^{j}_{mn}p_n\ln\frac{W^{j}_{mn}p_n}{W^{j}_{nm}p_m}.
\label{eq20}
\end{equation}
Due to periodicity of the steady-state, the integral
$\int_{0}^{\mathcal{T}}dt\,\sum_{m,n}W^{j}_{mn}p_n\ln (p_n/p_m)$ is strictly zero
and  Eq.~(\ref{eq20}) then reduces to
\begin{equation}
\bar {\sigma}^{j}=\frac{1}{\mathcal{T}}\sum_{m,n}\int_{0}^{\mathcal{T}}dt\,W^{j}_{mn}p_n\ln\frac{W^{j}_{mn}}{W^{j}_{nm}}.
\label{eq21}
\end{equation}
Since the ratio between $W^{j}_{mn}$ 
and $W^{j}_{nm}$ is given by the  local detailed balance condition, 
we can derive an expression for $\bar {\sigma}^{j}$ in terms
of thermodynamic variables;
\begin{eqnarray}
\bar {\sigma}^{j}&=&-\frac{1}{\mathcal{T}}\sum_{m}\int_{0}^{\mathcal{T}}dt\,\Big[\frac{\epsilon_m(t)-\mu_j(t)n_m}{T_j(t)} \Big]{\dot p^{j}}_m\nonumber\\&=&-\frac{1}{\mathcal{T}}\int_{0}^{\mathcal{T}}dt\,\frac{\dot{Q}_j(t)}{T_j(t)},
\end{eqnarray}
in agreement with the classical thermodynamic definition of entropy production \cite{callen1998thermodynamics}.

By inserting the expressions for heat and temperature from Eqs. (\ref{eq8}) and (\ref{eq14-3}), we have that 
\begin{eqnarray}
\bar {\sigma}=-\sum_j\frac{1}{\mathcal{T}}\int_{0}^{\mathcal{T}}dt\,\sum_{m}[\epsilon_m(t)-\mu_j(t)n_m]{\dot p^{j}_m(t)}\times\nonumber\\\Big[\frac{1}{T_0}+F_{T_j}g_{T_j}(t)\Big].
\label{eq19}
\end{eqnarray}
This suggests the introduction of a new thermodynamic flux,
\begin{equation}
J_{T_j}=-\frac{1}{\mathcal{T}}\int_{0}^{\mathcal{T}}dt\,{g_{T_j}(t)}\sum_{m,n}\gamma_{T_j,m}W_{mn}^{j}(t)p_n(t),
\label{eq19-1}
\end{equation}
which allows  us to rewrite the stochastic thermodynamics formula for entropy production to
a bilinear  function of thermodynamic forces and fluxes given by
\begin{equation}
{\bar \sigma}=\sum_j(F_{\epsilon_j} J_{\epsilon_j}+F_{\mu_j} J_{\mu_j}+F_{T_j} J_{T_j}),
\label{eq19-2}
\end{equation}
with $F_{\epsilon_j}=F_\epsilon$. This is in agreement with classical non-equilibrium thermodynamics \cite{de2013non}.

It is worth noting that the structure of Eq.~(\ref{eq19-1}) clearly mimics that of the work and chemical fluxes, Eqs.~(\ref{eq16-1}) and 
(\ref{Jepsi}). In fact one can easily verify that all three types of thermodynamic fluxes are of the form
\begin{equation}
J_{\alpha_j}=-\frac{1}{\mathcal{T}}\int_{0}^{\mathcal{T}}dt\,{g_{\alpha_j}(t)}\sum_{m,n}\gamma^{\alpha}_{m}W_{mn}^{j}(t)p_n(t),
\label{JGen}
\end{equation}
with $\alpha=T,\epsilon,\mu$.

\section{Onsager coefficients\label{ons}}
As the thermodynamic fluxes
vanish in the absence of thermodynamic forces, one expects that they depend
linearly on the thermodynamic forces $F_{\beta}$ near equilibrium, which implies the following form for  a flux $J_{\alpha}$:
\begin{equation}
J_{\alpha}=\sum_{\beta}{\it L}_{\alpha,\beta}F_{\beta},
\label{eq22}
\end{equation}
where ${\it L}_{\alpha,\beta}$ are the so-called Onsager coefficients.
From Eq.~(\ref{eq19-2}), the entropy production rate $\bar{\sigma}$ is depicted as a quadratic function of the thermodynamic forces,
\begin{equation}
    \bar{\sigma}=\sum_{\alpha,\beta}F_\alpha L_{\alpha,\beta}F_\beta.
\end{equation}

In the absence of odd parity variables (such as magnetic fields),
the Onsager coefficients of steady-state systems generally satisfy
the Onsager reciprocal relations, $L_{\alpha,\beta}=L_{\beta,\alpha}$. This is no longer the case for systems with time-dependent driving, since the
driving breaks the time-reversal symmetry. In this  instance, Onsager
symmetry is replaced by the weaker Onsager-Casimir symmetry, which relates the Onsager coefficients under time-forward
driving to the cross-coefficient of time-inverted driving,
\begin{equation}
  {\it L}_{\alpha,\beta}={\it \tilde{L}}_{\beta,\alpha},
\end{equation} 
where the tilde stands for time inverted driving,
$\tilde{g}_\alpha(t)=g_\alpha(-t)$.

Our aim here is to evaluate the Onsager coefficients 
and to prove the Onsager-Casimir reciprocal relations  for a periodically driven
system in contact with multiple reservoirs. By expanding
Eq.~(\ref{JGen})  up to first order in the thermodynamic forces,
we verify that $J_\alpha$ has two terms, 
one associated with the first order expansion of the transition matrix $W^j_{mn}$
(from Eqs.~(\ref{Walp-0}) and (\ref{Walp}))
and the other with the expansion of the probability
distribution $p_{n}(t)$ (from Eq.~(\ref{pj})). The total
flux  then reads  $J_{\alpha_j}=J^{(1)}_{\alpha_j}+J^{(2)}_{\alpha_j}$, with
\begin{equation}
J_{\alpha_j}^{(1)}=-\sum_{\beta}\Big[\frac{1}{\mathcal{T}}\int_{0}^{\mathcal{T}} dt\,g_{\alpha_j}(t) g_{\beta_j}(t) \sum_{m,n}\gamma_m^{\alpha}W_{mn}^{eq,j}\gamma_n^{\beta}p_n^{eq} \Big]F_{\beta_j},
\end{equation}
and
\begin{eqnarray}
J_{\alpha_j}^{(2)}=-\sum_{\beta,j^{'}} \Big[ \frac{1}{\mathcal{T}}\int_{0}^{\mathcal{T}}dt\,  \int_{0}^{\mathcal{\infty}} d\tau\, g_{\alpha_j}(t)g_{\beta_{j^{'}}}(t-\tau)\times\nonumber\\ 
\sum_{k,l,m,n} \gamma_m^{\alpha}W_{mn}^{eq,j}(e^{W^{eq}\tau})_{nk}W_{kl}^{eq,j^{'}}\gamma_{l}^{\beta}p_{l}^{eq}\Big]F_{\beta_{j^{'}}},
\end{eqnarray}
respectively. One can easily see that the first flux  depends solely on the thermodynamic forces associated with the same reservoir, while the second 
flux is dependent on all thermodynamic forces. Using the linearized expressions, one can now associate an Onsager matrix with each of those fluxes, $J_{\alpha_j}^{(1)}=\sum_{\beta}L^{(1)}_{\alpha_j,\beta_j}F_{\beta_j}$ and $J_{\alpha_j}^{(2)}=\sum_{\beta,j^{'}}L^{(2)}_{\alpha_j,\beta_{j^{'}}}F_{\beta_{j^{'}}}$, where the Onsager coefficients are given by
\begin{eqnarray}
L^{(1)}_{\alpha_j,\beta_j}=-\frac{1}{\mathcal{T}}\int_{0}^{\mathcal{T}}dt\, g_{\alpha_j}(t)g_{\beta_j}(t) \sum_{m,n}\gamma_m^{\alpha}W_{mn}^{eq,j}\gamma_n^{\beta}p_n^{eq},
\label{eq23-0}
\end{eqnarray}
and
\begin{eqnarray}
L^{(2)}_{\alpha_j,\beta_{j^{'}}}=-\frac{1}{\mathcal{T}}\int_{0}^{\mathcal{T}}dt\int_{0}^{\infty}d\tau\, g_{\alpha_j}(t)g_{\beta_{j^{'}}}(t-\tau) \times\nonumber\\
 \sum_{k,l,m,n} \gamma_m^{\alpha}W_{mn}^{eq,j}(e^{W^{eq}\tau})_{nk}W_{kl}^{eq,j^{'}}\gamma_{l}^{\beta}p_{l}^{eq},
\label{eq23-1}
\end{eqnarray}
with $L^{(1)}_{\alpha_j,\beta_{j'}}=0$ for $j\neq j'$. The total Onsager matrix is the sum of these two matrices, $L_{\alpha_j,\beta_{j'}}=L^{(1)}_{\alpha_j,\beta_{j'}}+L^{(2)}_{\alpha_j,\beta_{j'}}$. This structure for the Onsager coefficients resembles the one found for a class of quantum mechanical systems studied in \cite{brandner2016periodic}.

We are now ready to study the reciprocal relations for $L^{(1)}_{\alpha_j,\beta_j}$ and $L^{(2)}_{\alpha_j,\beta_{j^{'}}}$. Remarkably, both Onsager matrices will satisfy an Onsager-Casimir relation separately, which implies that the total Onsager matrix will satisfy the same Onsager-Casimir symmetry. We first look at $L^{(1)}_{\alpha_j,\beta_j}$. One can easily verify that these coefficients are invariant under time-reversal by replacing $g_\alpha(t)$ and $g_\beta(t)$ by $g_\alpha(-t)$ and $g_\beta(-t)$ and doing a change of integration variable to $t'=-t$. Subsequently,
 taking into account the detailed
balance condition, one can show that $\sum_{m,n}\gamma_n^{\beta}W_{nm}^{eq,j}\gamma_m^{\alpha}p_m^{eq}=\sum_{m,n}\gamma_m^{\alpha}W_{mn}^{eq,j}\gamma_n^{\beta}p_n^{eq}$ and
{ hence} the right side of Eq.~(\ref{eq23-0}) becomes
\begin{eqnarray}
L^{(1)}_{\alpha_j,\beta_j}&=&-\frac{1}{\mathcal{T}}\int_{0}^{\mathcal{T}}dt\, \tilde{g}_{\beta_j}(t)\tilde{g}_{\alpha_j}(t) \sum_{m,n}\gamma_n^{\beta}W_{nm}^{eq,j}\gamma_m^{\alpha}p_m^{eq}\nonumber\\&=&\tilde{L}^{(1)}_{\beta_j,\alpha_j},
\end{eqnarray}
which proves the Onsager-Casimir symmetry for $L^{(1)}_{\alpha_j,\beta_j}$.

The proof of the second term can be done in a similar way. We start by taking
the time  transformation $-t=t'-\tau$ in
Eq.~(\ref{eq23-1}),
\begin{eqnarray}
L^{(2)}_{\alpha_j,\beta_{j^{'}}}=\Big[\frac{1}{\mathcal{T}}\int_{0}^{\infty}d\tau\int_{\tau}^{\mathcal{-T+\tau}}dt'\, g_{\alpha_j}(-t'+\tau)g_{\beta_{j^{'}}}(-t')\nonumber\\
  \times \sum_{k,l,m,n} \gamma_m^{\alpha}W_{mn}^{eq,j}(e^{W^{eq}\tau})_{nk}W_{kl}^{eq,j^{'}}\gamma_{l}^{\beta}p_{l}^{eq} \Big].
\label{eq25}
\end{eqnarray}
Due to the periodicity of the drivings, the  first integral 
can be shifted
by $\mathcal{T}-\tau$, allowing us to rewrite it as
\begin{eqnarray}
L^{(2)}_{\alpha_j,\beta_{j^{'}}}=-\Big[\frac{1}{\mathcal{T}}\int_{0}^{\mathcal{T}}dt'\int_{0}^{\infty}d\tau\,  {\tilde g}_{\alpha_j}(t'-\tau) {\tilde g}_{\beta_{j^{'}}}(t')\nonumber\\
\times  \sum_{k,l,m,n} \gamma_m^{\alpha}W_{mn}^{eq,j}(e^{W^{eq}\tau})_{nk}W_{kl}^{eq,j^{'}}\gamma_{l}^{\beta}p_{l}^{eq} \Big].
\label{eq26} 
\end{eqnarray}
By once again appealing to the detailed balance condition, one can show that
\begin{eqnarray}&\sum_{k,l,m,n} \gamma_m^{\alpha}W_{mn}^{eq,j}(e^{W^{eq}\tau})_{nk}W_{kl}^{eq,j^{'}}\gamma_{l}^{\beta}p_{l}^{eq}\nonumber\\&=\sum_{k,l,m,n} \gamma_{l}^{\beta}W_{lk}^{eq,j^{'}}(e^{W^{eq}\tau})_{kn}W_{nm}^{eq,j}\gamma_m^{\alpha}p_{m}^{eq},\end{eqnarray} 
and hence $L^{(2)}_{\alpha_j,\beta_{j^{'}}}$ becomes
\begin{eqnarray}
L^{(2)}_{\alpha_j,\beta_{j^{'}}}=-\Big[\frac{1}{\mathcal{T}}\int_{0}^{\mathcal{T}}dt'\int_{0}^{\infty} d\tau {\tilde g}_{\beta_{j^{'}}}(t'){\tilde g}_{\alpha_j}(t'-\tau) \nonumber\\
\times \sum_{k,l,m,n} \gamma_{l}^{\beta}W_{lk}^{eq,j^{'}}(e^{W^{eq}\tau})_{kn}W_{nm}^{eq,j}\gamma_m^{\alpha}p_{m}^{eq} \Big].
\label{eq27} 
\end{eqnarray}
The right hand side of Eq.~(\ref{eq27}) is just the Onsager coefficient ${\tilde L}_{\beta_{j^{'}},\alpha_j}$. This completes the proof of the Onsager-Casimir symmetry.

These results also implies an Onsager-Casimir relation for any combination of $J_{\alpha,i}$'s. For example, if we define,
\begin{equation}
    J'_i=\sum_{\alpha,j}A_{i;\alpha,j}J_{\alpha,j},
\end{equation}
for some matrix $A$, then the associated thermodynamic forces are given by
\begin{equation}
    F'_i=\sum_{\alpha,j}\left(A^{-1}\right)_{\alpha,j;i}J_{\alpha,j},
\end{equation}
as the entropy production rate $\sigma=\sum_iF'_iJ'_i$ is independent of the choice of $J$'s. The new Onsager matrix $L'$ is of the form,
\begin{equation}
    L'=ALA^\dag,
\end{equation}
and one can straightforwardly verify that this matrix should satisfy the same Onsager-Casimir relations as the original matrix $L$. In particular, we can conclude that Onsager-Casimir symmetry is also valid when one only looks at the total fluxes $J_\epsilon$, $J_\mu$ and $J_T$.

\section{Two level systems \label{app}}
\begin{figure*}
  \centering
\includegraphics[scale=0.6]{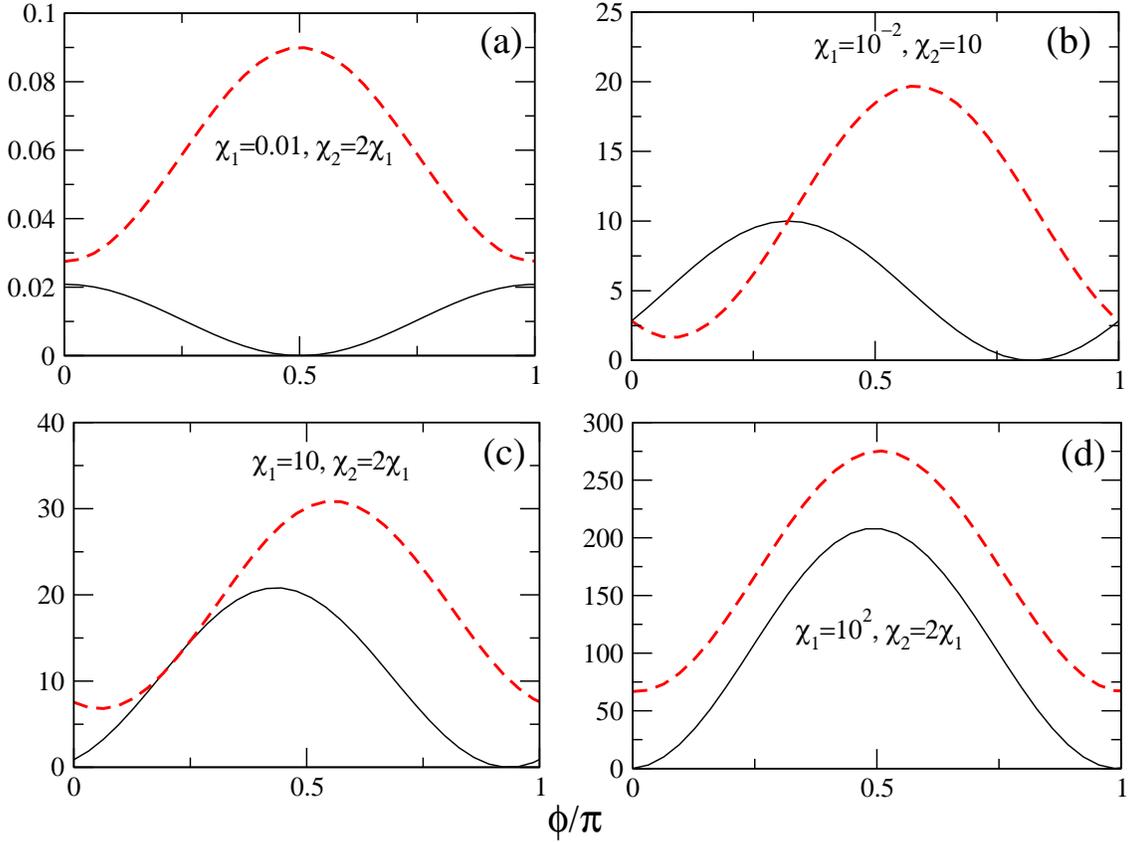}
\caption{ Reduced dissipated power -$\dot {\bar W}\mathcal{T}/(T_0p^{eq}_0p^{eq}_1F_{T_1}^2)$ (continuous lines) and
  entropy production ${\bar \sigma}\mathcal{T}/(p^{eq}_0p^{eq}_1F_{T_1}^2)$ (dashed
  lines) versus $\phi$ for distinct values of couplings $\chi_1$ and $\chi_2$.
  In all cases we considered $F_{T_2}=2F_{T_1}$.  }
\label{fig1}
\end{figure*}
As an example, we consider a
quantum dot, with one active energy level, in contact
with two electron reservoirs at temperatures $T_1(t)$ and $T_2(t)$ and chemical potentials $\mu_1(t)$ and $\mu_2(t)$, respectively. The quantum dot can be empty or occupied 
by a single electron  with probabilities $p_0(t)$ and $p_1(t)=1-p_0(t)$,
respectively.
The temperature and chemical of the electron  reservoirs 
as well as the energy of the quantum dot are modulated
according to Eqs.~(\ref{eq8})-(\ref{eq7}), respectively.
The total transition matrix ${W(t)}$ is the sum of both reservoir
contributions
${W(t)}={W^1(t)+W^{(2)}(t)}$, with $W^j(t)$  given by
\begin{displaymath}
{W^j(t)}=\left(\begin{array}{cc}
  -{\Gamma_j y(t)} & {\Gamma_j (1-y(t))}\\
 {\Gamma_j y(t)} & -{\Gamma_j (1-y(t))}
\end{array}\right),
\end{displaymath}
where $\Gamma_j$ describes the interaction between the quantum dot
and the $j$-th
reservoir and $y(t)$ is the Fermi-Dirac distribution 
$y(t)=[1+\exp((\epsilon(t)-\mu(t))/T(t))]^{-1}$.

For simplicity, we set $F_{\mu_j}=0$, implying that 
the chemical potentials of both reservoirs are the same
and thereby $\dot{\bar{W}}_{chem}=0$. { The thermodynamic variables are modulated} via the driving functions
 $g_{\epsilon}(t)=\sqrt{2}\sin(\omega t+\phi)$ and
$g_{T_j}(t)=\sqrt{2}\sin(\omega t)$,
respectively, where we have assumed that both reservoirs and the electron level are all driven with the same frequency $\omega=2\pi/\mathcal{T}$, but with a phase-difference $\phi$  between the driving of the energy and the
and temperature drivings.

The lowest order expressions for the energy and temperature fluxes $J_\epsilon$ and $J_{T_{j}}$ and for the entropy production rate  ${\bar \sigma}$ can now be calculated using the expressions from the previous section  and are given by
\begin{widetext}
\begin{equation}
J_{\epsilon}=\frac{2\pi p^{eq}_0p^{eq}_1}{\mathcal{T}(4\pi^2+{\tilde \chi}^{2})}[2\pi F_{\epsilon}{\tilde \chi}+(\epsilon_0-\mu_0)({\tilde \chi}\sin\phi+2\pi\cos\phi)](F_{T_1}\chi_1+F_{T_2}\chi_2)],
\end{equation}
\begin{equation}
  J_{T_1}=\frac{p^{eq}_0p^{eq}_1(\epsilon_0-\mu_0)\chi_1}{\mathcal{T}(4\pi^{2}+{\tilde \chi}^{2})}[F_{\epsilon}(4\pi^2\cos\phi-2\pi{\tilde \chi}\sin\phi)+(\epsilon_0-\mu_0){\tilde \chi}\chi_2 (F_{T_1}-F_{T_2})+4\pi^2(\epsilon_0-\mu_0)F_{T_1}],
  \end{equation}
  \begin{equation}
  {\bar \sigma}=\frac{4\pi^2 p^{eq}_0p^{eq}_1}{\mathcal{T}(4\pi^{2}+{\tilde \chi}^{2})}[F^{2}_\epsilon{\tilde \chi}+(\epsilon_0-\mu_0)^{2}(F^{2}_{T_1}\chi_1+F^{2}_{T_2}\chi_2)+
2(\epsilon_0-\mu_0)\cos\phi F_\epsilon(F_{T_1}\chi_1+F_{T_2}\chi_2)+
\frac{(\epsilon_0-\mu_0)^{2}}{4\pi^2}\chi_1\chi_2{\tilde \chi}(F_{T_1}-F_{T_2})^2 ],
\end{equation}
\end{widetext}
respectively, where $\chi_j=\Gamma_j\mathcal{T}$ and ${\tilde \chi}=(\Gamma_1+\Gamma_2)\mathcal{T}$.
{For the special case, $F_{T_1}=F_{T_2}$ and
  $\Gamma_1=\Gamma_2$, the above expressions reduce to the single reservoir case up to a factor $2$ \cite{proesmans2016linear}. 
$J_{T_2}$ has a similar expression as $J_{T_1}$ with $T_1$ and $T_2$, 
$F_1$ and $F_2$ interchanged.} One can easily verify that these expressions satisfy Onsager-Casimir symmetry.

These results can be used to optimize the amount of mechanical work, $\dot {\bar W}=T_0F_{\epsilon}J_\epsilon$ that can be done by the system. 
In particular, an optimization with respect to $F_{\epsilon}$ yields
 the following relation for the maximum mechanical power:
\begin{multline}
    -\dot {\bar W}_{\mathrm{max}}=\frac{p^{eq}_0p^{eq}_1T_0(\epsilon_0-\mu_0)^2(F_{T_1}\chi_1+F_{T_2}\chi_2)^2}{{\bf 4}{\tilde \chi}\mathcal{T}(4\pi^{2}+{\tilde \chi}^{2})}\\\times({\tilde \chi}\sin\phi+2\pi\cos\phi)^2.
\label{eq29}
\end{multline}
Fig.~\ref{fig1} depicts the behavior
of $\sigma$ and $\dot {\bar W}$ versus $\phi$
for distinct  couplings $\chi_1$ and $\chi_2$ with $F_\epsilon$
given by its optimal value.
In the
limit of low (large) ``effective''
couplings, ${\tilde \chi} \ll 1$ (${\tilde \chi} \gg 1$),
the work output is maximal (minimum) when the driving of the work source is in phase with that of the heat sources, $\phi=0$ and minimum (maximum) mwhen the driving is out of phase, $\phi=\pi/2$,
in accordance
with  Eq.~(\ref{eq29}). Conversely, for the above
choice of $F_\epsilon$, the positions of maxima and minima of
 the entropy
production fulfill the above relation
\begin{eqnarray}
\tan\phi=\left\{-\frac{{\tilde \chi}^2+12\pi^2\pm\sqrt{{\tilde \chi}^4+40{\tilde \chi}^2\pi^2+144\pi^4}}{4\pi{\tilde\chi}} \right\},
\end{eqnarray}
where $+(-)$ denote to the maximum (minimum).
They approach $\pi/2$ (maximum) 
and $\phi=0$ (minimum) for 
${\tilde \chi} \ll 1$ and ${\tilde \chi} \gg 1$, respectively and deviate
from these limits for intermediate coupling sets.

Similar analytic optimizations for other
thermodynamic fluxes can also be performed. 
For instance, by optimizing  $\dot {\bar W}$
with respect to both $F_\epsilon$ and
the phase-difference $\phi$, the expression for $\dot {\bar W}_{\mathrm{max}}$
becomes
\begin{equation}
    -\dot {\bar W}_{\mathrm{max}}=\frac{p^{eq}_0p^{eq}_1T_0(\epsilon_0-\mu_0)^2(F_{T_1}\chi_1+F_{T_2}\chi_2)^2}{{ 4}{\tilde \chi}\mathcal{T}},
\end{equation}
where the optimal phase-difference and amplitude are given by
\begin{eqnarray}
\tan\phi&=&\left(\frac{\tilde \chi}{2\pi}\right)\\
  F_\epsilon&=&-\frac{(\epsilon_0-\mu_0)}{4\pi {\tilde \chi}}\sqrt{4\pi^{2}+{\tilde \chi}^{2}}(F_{T_1}\chi_1+F_{T_2}\chi_2).\,\;
\label{eq30}
\end{eqnarray}

\section{Conclusions\label{con}}
In this paper, we have derived a general linear description 
for the thermodynamics of Markov systems in contact to multiple reservoirs, using the framework of stochastic thermodynamics.
We have shown that the thermodynamic fluxes such as mechanical and chemical work and heat, can be written in a general form, as functions of the driving and the equilibrium properties of the system. The entropy production is obtained as a bilinear function of thermodynamic 
forces and associated fluxes. Furthermore, we calculated all Onsager coefficients and showed that they satisfy a generalized Onsager-Casimir relation.

Finally, we mention some interesting directions for 
further research. Firstly, it would be interesting to  extend our analysis to higher order response coefficients and to study the resulting constraints on heat engines \cite{cleuren2015universality,vroylandt2018degree}. 
Secondly, it would be very interesting to see if our results can be extended to quantum mechanical systems and systems with strong coupling \cite{vinjanampathy2016quantum,jarzynski2017stochastic}. 
Finally, it should be no problem to test our predictions, such as the generalized Onsager-Casimir relation, with state-of-the-art experimental setups \cite{pekola2015towards,ciliberto2017experiments}.

\begin{acknowledgements}
We thank Bart Cleuren for a careful reading of the manuscript.
KP is a postdoctoral fellow of the Research Foundation-Flanders (FWO). 	C.E.F.  acknowledges the financial support from FAPESP		
under grant No 2018/02405-1
\end{acknowledgements}








\end{document}